\documentclass{birkjour_t2}
\usepackage{amssymb,amsmath,amsthm,amscd,amsrefs,graphicx}
\usepackage{hyperref}
\hypersetup{
  colorlinks   = true,	%Colour links instead of ugly boxes
  urlcolor     = blue,	%Colour for external hyperlinks
  linkcolor    = blue,	%Colour of internal links
  citecolor   = red	%Colour of citations
}
\newcommand\scalemath[2]{\scalebox{#1}{\mbox{\ensuremath{\displaystyle #2}}}}
\newcommand{\hpg}[5]{{}_{#1}F_{#2}\! \left(\left.{\genfrac{}{}{0pt}{}{#3}{#4}}\right| #5 \right) }
\newcommand{\app}[4]{F_{#1}\! \left(\left.{\genfrac{}{}{0pt}{}{#2}{#3}}\right| #4 \right) }
\title[Parabolic Vortex Segments using Elliptic Integrals]{Application of the Biot-Savart Law to Parabolic Vortex Segments using Elliptic Integrals}
\author{Andreas Malmendier}
\address{Department of Mathematics and Statistics, Utah State University, Logan, UT 84322}
\email{andreas.malmendier@usu.edu}
\thanks{A.M. acknowledges support from the Simons Foundation through grant no.~202367.}

\author{Jackson T. Reid}
\address{Department of Mechanical and Aerospace Engineering, Utah State University, Logan, Utah, 84321, USA}
\begin{document}
\begin{abstract}
The Biot-Savart law is used in aerodynamic theory to calculate the velocity induced by curved vortex lines.  
Explicit formulas are developed, using multivariate Appell hypergeometric functions, for the velocity induced by a general parabolic vortex segment. 
The formulas are derived by constructing a particular pencil of elliptic curves whose period integrals provide the 
solution to the induced velocity. We use numerical integration and a perturbation expansion to evaluate the validity of our formulas.
\end{abstract}
\maketitle
\section{Introduction}
In the field of aerodynamics, potential flow theory is often used to predict aerodynamic performance and characteristics~\cites{Complex_text, Fluids_text, Aero_Design, MechFlight, Aerodynamics}. Algorithms like vortex panel methods, vortex lattice methods, and lifting-line theory use combinations of potential-flow elements (vortices, sources and sinks, etc.) to model the flow around bodies~\cites{Swept_ZeroLift, Wing_Loading}.  Many of these methods use linear potential-flow elements to approximate the non-linear strength distributions or geometries of the elements in the flow~\cites{Swept_Lift, Num_Lifting_Line, horseshoe_sheet, EVSM, wood2001accurately,gupta2005accuracy}. The ability to apply higher-order, non-linear potential-flow elements in these algorithms achieves higher accuracy for a fewer number of elements, potentially reducing the computational cost of the method~\cites{montgomery2018propeller, govindarajan2015curvature, bliss1987new, van2012core, bhagwat2014self, kim2016dynamic, yoon2004analytical, beyer2012development, nagati1987vortex}. 

Work to predict the influence of various curved vortex segments has been done before. In \cites{yoon2004analytical} analytic predictions for the influence of a thin vortex ring were developed. The authors of \cites{beyer2012development} developed a closed-form prediction for the influence of a circular-arc vortex segment with a cut-off radius. In \cites{bliss1987new} it was also predicted the self-influence of a circular-arc vortex segment, and predicted the influence of a symmetric parabolic vortex segment using simplifying approximations. The work performed herein produces a closed-form prediction of the influence of a non-symmetric parabolic vortex segment in three-dimensions, allowing for more applicability than circular-arc segments, without the approximations made in \cites{bliss1987new}.

\subsection{Parabolic Vortex Segment}
The Biot-Savart law is used in aerodynamic theory to calculate the velocity induced by vortex lines. The Biot-Savart law describes the differential velocity, $d\boldsymbol{V}$, induced by the differential element, $d\boldsymbol{l}$, of a vortex filament in three dimensions
\begin{equation}\label{e:Biot_Savart1}
	d\boldsymbol{V} = \frac{\Gamma}{4 \pi} \frac{ d\boldsymbol{l} \times \boldsymbol{r}}{ \lvert \boldsymbol{r} \rvert^3} \,,
\end{equation}
where $\Gamma$ is the strength of the vortex, and $\boldsymbol{r}$ is the position vector from the differential vortex segment to the point, $\boldsymbol{x}$, at which the differential induced velocity is calculated~\cites{MechFlight}. If the vortex follows the parameterized curve, $\boldsymbol{f}(t)$, then equation~\eqref{e:Biot_Savart1} can be rewritten as
\begin{equation}\label{e:Biot_Savart2}
	d\boldsymbol{V} = \frac{\Gamma}{4 \pi} \frac{ d\boldsymbol{f}(t) \times \big(\boldsymbol{x} - \boldsymbol{f}(t) \big)}{ \lvert \boldsymbol{x} - \boldsymbol{f}(t) \rvert^3} \,,
\end{equation}
where $d\boldsymbol{f}(t)$ is the derivative of $\boldsymbol{f}(t)$ with respect to the parameter $t$.

\begin{figure}
	\centering
	\includegraphics{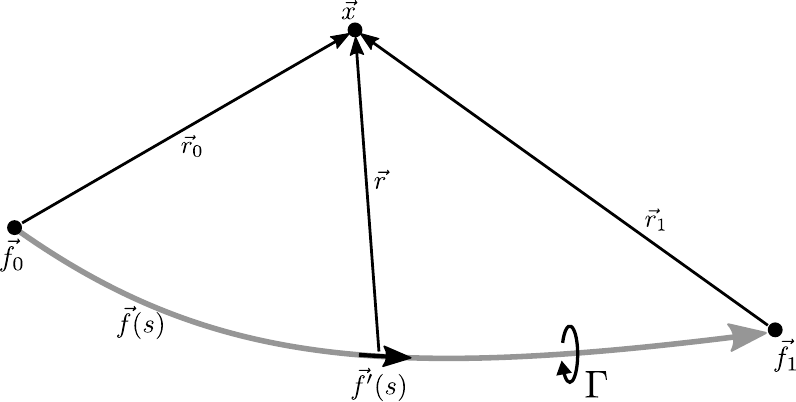}
	\caption{Diagram of the velocity induced by a conic vortex segment at the point $\boldsymbol{x}$.}
	\label{f:conic_segment}
\end{figure}

Consider a parabolic vortex segment beginning at point $\boldsymbol{f}_0$ and ending at point $\boldsymbol{f}_1$, as shown in figure~\ref{f:conic_segment}, and let it be defined by the parameterized curve
\begin{align}\label{e:conic_segment_function}
	\boldsymbol{f}(t) &= (\boldsymbol{r}_0 - \boldsymbol{r}_1 - \boldsymbol{f}_0') t^2 + \boldsymbol{f}_0't + \boldsymbol{f}_0 \nonumber \\
	d\boldsymbol{f}(t) &= \big(2(\boldsymbol{r}_0 - \boldsymbol{r}_1 - \boldsymbol{f}_0')t + \boldsymbol{f}_0' \big)dt 
\end{align}
for $0 \leq t \leq 1$. We remark that the quantity $\boldsymbol{r}_0 - \boldsymbol{r}_1 - \boldsymbol{f}_0'$ is meant to represent the constant second derivative vector of the parabolic segment $\boldsymbol{f}(t)$. As we normalized $\boldsymbol{r}_1$ to represent a vector at time $t=1 [\text{time}]$, it would formally correct to write $(\boldsymbol{r}_0 - \boldsymbol{r}_1)[\text{time}]^{-2} - \boldsymbol{f}_0'[\text{time}]^{-1}$. In the following we will simply drop the symbol $[\text{time}]$.
Using equation~\eqref{e:conic_segment_function} in equation~\eqref{e:Biot_Savart2}, results in the expression
\begin{equation}
	d\boldsymbol{V} = \frac{\Gamma}{4 \pi} \frac{ \big(2(\boldsymbol{r}_0 - \boldsymbol{r}_1 - \boldsymbol{f}_0')t + \boldsymbol{f}_0' \big)dt \times \big(\boldsymbol{x} - (\boldsymbol{r}_0 - \boldsymbol{r}_1 - \boldsymbol{f}_0') t^2 - \boldsymbol{f}_0't - \boldsymbol{f}_0 \big)}{ \lvert \boldsymbol{x} - (\boldsymbol{r}_0 - \boldsymbol{r}_1 - \boldsymbol{f}_0') t^2 - \boldsymbol{f}_0't - \boldsymbol{f}_0 \rvert^3} \,,
\end{equation}
which can be rewritten using the definitions shown in figure~\ref{f:conic_segment} as
\begin{equation}\label{e:conic_segment_dV}
\scalemath{0.8}{	\frac{d\boldsymbol{V}}{dt} = \frac{\Gamma}{4 \pi} \frac{\big( \boldsymbol{f}_0' \times (\boldsymbol{r}_0 - \boldsymbol{r}_1) \big)t^2 - 2 \big( (\boldsymbol{r}_1 + \boldsymbol{f}_0') \times \boldsymbol{r}_0\big)t  +  \boldsymbol{f}_0' \times \boldsymbol{r}_0 }{\Big(  (\boldsymbol{r}_0 - \boldsymbol{r}_1 - \boldsymbol{f}_0')^2 t^4 + 2(\boldsymbol{r}_0 - \boldsymbol{r}_1 - \boldsymbol{f}_0') \cdot \boldsymbol{f}_0' t^3 + \big( f_0'^2 - 2\boldsymbol{r}_0 \cdot (\boldsymbol{r}_0 - \boldsymbol{r}_1 - \boldsymbol{f}_0') \big)t^2 - 2\boldsymbol{r}_0 \cdot \boldsymbol{f}_0' t +  r_0^2 \Big)^{\frac{3}{2}} }} \,,
\end{equation}
where $r_0 = \lvert \boldsymbol{r}_0 \rvert$, $r_1 = \lvert \boldsymbol{r}_1 \rvert$, and $f_0' = \lvert \boldsymbol{f}_0' \rvert$. The total velocity induced at $\boldsymbol{x}$ by the vortex segment is calculated as equation~\eqref{e:conic_segment_dV} is integrated along the length of the vortex segment, resulting in an integral of the form
\begin{equation}\label{e:parabolic_integral_form}
	\boldsymbol{V} = \frac{\Gamma}{4 \pi} \int_0^{1} \frac{\Big( A t^2 + B t  +  C \Big) dt}{\Big( D t^4 + E t^3 + F t^2 + G t +  H \Big)^{\frac{3}{2}} } \,,
\end{equation}
where
\begin{equation*}
	A = \boldsymbol{f}_0' \times (\boldsymbol{r}_0 - \boldsymbol{r}_1),\quad B = -2(\boldsymbol{r}_1 + \boldsymbol{f}_0') \times \boldsymbol{r}_0,\quad C = \boldsymbol{f}_0' \times \boldsymbol{r}_0,
\end{equation*}
\begin{equation*}
	D = (\boldsymbol{r}_0 - \boldsymbol{r}_1 - \boldsymbol{f}_0')^2 ,\quad E = 2(\boldsymbol{r}_0 - \boldsymbol{r}_1 - \boldsymbol{f}_0') \cdot \boldsymbol{f}_0',\quad F = f_0'^2 - 2\boldsymbol{r}_0 \cdot (\boldsymbol{r}_0 - \boldsymbol{r}_1 - \boldsymbol{f}_0'),
\end{equation*}
\begin{equation*}
	G = -2\boldsymbol{r}_0 \cdot \boldsymbol{f}_0' ,\quad H = r_0^2 \,.
\end{equation*}
equation~\eqref{e:parabolic_integral_form} belongs to the family of elliptical integrals, for which special integration techniques, described herein, are required to obtain a general, closed-form solution.

\subsection{Previous Work}
Before the solution to the general, parabolic vortex segment is discussed, it is important to consider the work previously completed in the prediction of vortex segment influence. This previous work provides a means against which to verify the fidelity of mathematical derivations and the accuracy of numerical results. First, the simplifying case of a linear vortex segment is described. Then, the work of \cites{bliss1987new} is summarized. Substantial work in the field of circular arc vortex segments has been performed~\cites{montgomery2018propeller, govindarajan2015curvature, bliss1987new, van2012core, bhagwat2014self, kim2016dynamic, yoon2004analytical, beyer2012development}. While a circular arc may provide a comparison for the case of a parabolic vortex in the limit of a small arc length, it will not be described herein, but will be considered in future work.

\subsubsection{Linear Vortex Segment}\label{s:Linear_segment}
In the special case that $\boldsymbol{f}'_0 =\boldsymbol{r}_0 - \boldsymbol{r}_1$, the parabolic vortex segment becomes a line that extends from the point $\boldsymbol{f}_0$ to the point $\boldsymbol{f}_1$. Thus, for this linear vortex segment,
\begin{align}\label{e:linear_segment_function}
	\boldsymbol{f}(t) &=  \boldsymbol{f}'_0 t + \boldsymbol{f}_0 \nonumber \,,\\
	d\boldsymbol{f}(t) &= \boldsymbol{f}'_0 dt
\end{align}
for $0 \leq t \leq 1 \nonumber$. Using equation~\eqref{e:linear_segment_function} en lieu of equation~\eqref{e:conic_segment_function}, equation~\eqref{e:parabolic_integral_form} becomes
\begin{equation}\label{e:linear_integral_form}
	\boldsymbol{V} = \frac{\Gamma}{4 \pi} \int_0^{1} \frac{ \big( \boldsymbol{r}_0  \times \boldsymbol{r}_1 \big) dt }{ \Big( \big( r_0^2 + r_1^2- 2(\boldsymbol{r}_0 \cdot \boldsymbol{r}_1)  \big) t^2 - 2\big( r_0^2 - \boldsymbol{r}_0 \cdot \boldsymbol{r}_1 \big) t + r_0^2 \Big)^{\frac{3}{2}}}  \,,
\end{equation}
which can be evaluated to yield
\begin{equation}\label{e:linear_segment_V}
	\boldsymbol{V} = \frac{\Gamma \,\big( \boldsymbol{r}_0  \times \boldsymbol{r}_1 \big)}{4 \pi} \Bigg( \frac{ r_0 + r_1 }{ \big( \boldsymbol{r}_0 \cdot \boldsymbol{r}_1 + r_0 r_1 \big) r_0 r_1} \Bigg) \,.
\end{equation}

The simplicity of the closed-form solution described in equation~\eqref{e:linear_segment_V} allows it to readily be used. equation~\eqref{e:linear_segment_V} also provides a check for the general solution for the influence of a parabolic vortex segment obtained in this work. A correct solution to equation~\eqref{e:parabolic_integral_form} will return to equation~\eqref{e:linear_segment_V} in the case that $\boldsymbol{f}'_0 =\boldsymbol{r}_0 - \boldsymbol{r}_1$ .

\subsubsection{Previous Approximation of Parabolic Vortex Segment}\label{s:Bliss_method}
The influence of a parabolic vortex segment has been approximated by \cites{bliss1987new}, for the parabolic arc described by the equation
\begin{equation}\label{e:symmetric_parabolic_vortex_function}
	y = \varepsilon x^2 \,,
\end{equation}
where $\varepsilon$ is defined in terms of the radius of the circular arc, $R_0$, and the arc's central angle, $\theta_0$,
\begin{equation}\label{e:vareps}
	\varepsilon = \frac{\cos \frac{\theta_0}{2} - 1}{R_0 \sin^2 \frac{\theta_0}{2}} \,,
\end{equation}
and $x$ varies from $-\ell$ to $\ell$. With this definition for the arc, the differential arc-length vector can be written
\begin{equation}
	d\boldsymbol{l} = \big[ 1, 2\varepsilon x, 0 \big] \, dx \,,
\end{equation}
and the distance from a point on the arc to the point at which the influence is calculated, $\big[ x_0, y_0, z_0 \big]$, is
\begin{equation}
	\boldsymbol{r} = \big[ x_0 - x, y_0 - \varepsilon x^2, z_0 \big] \,.
\end{equation}
The authors of \cites{bliss1987new} then describe the influence of the arc at a point, using the Biot-Savart law, as
\begin{align}\label{e:Bliss_Biot_Savart1}
	\boldsymbol{V} &= \frac{\Gamma}{4 \pi} \int_C \frac{ d\boldsymbol{l} \times \boldsymbol{r}}{ \lvert \boldsymbol{r} \rvert^3} \nonumber \\
		   &= \frac{\Gamma}{4 \pi} \int_{-\ell}^{\ell} \frac{ \big[ 2\varepsilon z_0 x,\  -z_0, \ -2\varepsilon x x_0 + \varepsilon x^2 + y_0 \big]}{ \Big( \varepsilon^2 x^4 + (1 - 2\varepsilon y_0)x^2 + (-2x_0) x + (x_0^2 + y_0^2 + z_0^2) \Big)^{3/2}} dx \,.
\end{align}

In order to evaluate equation~\eqref{e:Bliss_Biot_Savart1}, the authors of \cites{bliss1987new} model the quartic term, $\varepsilon^2 x^4$, by the quadratic
\begin{equation}
	\varepsilon^2 x^4 \approx \varepsilon^2 \big( F_2(x_0) x^2 + F_1(x_0) x + F_0(x_0) \big) \,,
\end{equation}
where $F_2$, $F_1$, and $F_0$ are functions of $x_0$. Thus,
\begin{equation}\label{e:symmetric_parabolic_integral}
\scalemath{0.9}{	\boldsymbol{V} = \frac{\Gamma}{4 \pi} \int_{-\ell}^{\ell} \frac{ \big[ 2\varepsilon z_0 x, \ -z_0, \ -2\varepsilon x x_0 + \varepsilon x^2 + y_0 \big]}{ \Big( (1 - 2\varepsilon y_0 + \varepsilon^2 F_2)x^2 + (-2x_0 + \varepsilon^2 F_1)x + (x_0^2 + y_0^2 + z_0^2 + \varepsilon^2 F_0) \Big)^{3/2}} dx} \,.
\end{equation}
Evaluating this simpler integral yields an expression in which $F_2$, $F_1$, and $F_0$ that must be tuned, depending on $x_0$ and $\lvert \boldsymbol{r} \rvert$, to accurately replicate the original integral. In \cites{bliss1987new} the authors describe their tuning methodology in detail, and it will not be described herein.

Equation~\eqref{e:symmetric_parabolic_vortex_function} specifies a vortex segment that is equivalent to the special case of equation~\eqref{e:conic_segment_function}, in which
\begin{equation}
	\boldsymbol{f}_0 = \big[ -\ell, \varepsilon \ell^2, 0\big] \,,
\end{equation}
\begin{equation}
	\boldsymbol{r}_0 - \boldsymbol{r}_1 = \big[ 2 \ell, 0, 0 \big] \,,
\end{equation}
\begin{equation}
	\boldsymbol{f}'_0 = \big[ 2 \ell, -4\varepsilon \ell^2 , 0 \big] \,.
\end{equation}
Therefore, for this case, the general solution for the influence of a parabolic vortex segment may be compared against the results of equation~\eqref{e:symmetric_parabolic_integral}, as well as a numerical evaluation of equation~\eqref{e:parabolic_integral_form}. Such a comparison will provide insight into the accuracy of the general solution and the approximate solution made in \cites{bliss1987new}.
\section{Evaluation of the Induced Velocity using Elliptic Integrals}\label{s:Integral}
The evaluation of an integral of the form
\begin{equation}\label{eqn:original_integral}
	\int  \frac{A t^2 + B t  +  C  }{\Big( D t^4 + E t^3 + F t^2 + G t +  H \Big)^{\frac{3}{2}} }dt
\end{equation}
requires techniques beyond those used in traditional integration. The original integral is rewritten in terms of symmetric integrals, which are then redefined based on the relation between genus-one curves and their Jacobian varieties, and finally related to hypergeometric functions. Note that, for generality, the constant multiplier in equation~\eqref{e:parabolic_integral_form}, $\Gamma / 4 \pi$, has been factored out of equation~\eqref{eqn:original_integral} and does not appear throughout this derivation. It is re-included for the discussion of this work's application and results.

\subsection{Reduction to a Standard Symmetric Integral}
Consider, first, a genus-one curve, $\mathcal{C}$, given as a quartic curve with affine coordinates $(t,s) \in \mathbb{C}^2$ by
\begin{equation}
	\mathcal{C}: \quad s^2 = \prod_{i=1}^4 \big(\alpha_i + \beta_i t\big) = D t^4 + E  t^3 + F t^2 + G t +H \,,
\end{equation}
where it is always assumed that $s^2$ only has simple zeros in $t$. Let $R(s,t)$ be a rational function of $s$ and $t$, containing at least one odd power of $s$. Expressions of the form
\begin{equation}
	\int R(s,t) \, dt
\end{equation}
are called \textit{elliptic integrals}. Because $s^2$ is a polynomial in $t$, one can carry out a partial fraction decomposition and write
\begin{equation}
	R(s,t) = \frac{\rho}{s} + \sigma \,,
\end{equation}
where $\rho$ and $\sigma$ are functions of $t$ alone. Herein, the case where $\sigma = 0$ is of interest. In fact, the elliptic integral in equation~\eqref{eqn:original_integral}, is of the general form
\begin{equation}\label{eqn:start}
	\mathcal{F} = \int_Y^X \frac{p_2(t)}{s^2} \frac{dt}{s} = \int_Y^X \frac{A t^2 + B t  +  C  }{\Big( D t^4 + E t^3 + F t^2 + G t +  H \Big)^{\frac{3}{2}} }dt \,.
\end{equation}
Assume that the limits of integration are real (i.e. $X,Y \in \mathbb{R}$) and that for $1 \leq i \leq 4$ the line segments with endpoints $\alpha_i + \beta_i X$ and $\alpha_i + \beta_i Y$ lie entirely within the complex plane cut along the negative real axis; these conditions eliminate any potential ambiguity in the definition of the integral which can then be extended by complex analytic continuation.

The transformation $\Psi$, between the variables $(u,v)$ and $(t,s)$, given by
\begin{equation}\label{eqn:transfo}
	\Psi : \quad  z_i = \frac{\alpha_i + \beta_i Y}{\alpha_i + \beta_i X}\,, \quad u = -\frac{t - Y}{t - X}\,, \quad v = \frac{(X - Y)^2 s}{s_0(X - t)^2} \,,
\end{equation}
where $s_0 := s(X) = \sqrt{(\alpha_1 + \beta_1 X) \cdots (\alpha_4 + \beta_4 X)}$, transforms the genus one curve $\mathcal{C}$ into the normalized quartic equation with affine coordinates $(u,v) \in \mathbb{C}^2$, given by
\begin{equation}\label{eqn:genus_one}
	\mathcal{C}: \quad v^2 = \big(z_1 + u\big) \big(z_2 + u\big) \big(z_3 + u\big) \big(z_4 + u\big) \,.
\end{equation}
It is easy to check that the transformation described in equation~\eqref{eqn:transfo} relates the holomorphic differentials (via pull back),
\begin{equation}
	\frac{dt}{s} = \Psi^*\left(\frac{X-Y}{s_0} \frac{du}{v}\right) \,,
\end{equation}
as well as meromorphic differentials,
\begin{equation}
	\frac{p_2(t)}{s^2} \frac{dt}{s} = \Psi^*\left(\frac{X-Y}{s_0^3} \frac{du}{v} \frac{(1+u)^2\big(A'(1+u)^2 + B'(1+u) + C'\big)}{v^2}\right) \,,
\end{equation}
where
\begin{align}
	A' &= p_2(X) = AX^2 + BX + C \,,\nonumber \\
	B' &= -p'_2(X)(X-Y) = -(2AX + B)(X-Y) \,,\\
	C' &= \frac{1}{2} p_2''(X)(X-Y)^2 = A(X-Y)^2 \,.\nonumber
\end{align}
Thus, we obtain for the integral $\mathcal{F}$ in terms of the new variables $u, v$, related by equation~\eqref{eqn:genus_one}, the following expressions
\begin{equation}\label{eqn:int0}
	\mathcal{F} = \int_Y^X \frac{p_2(t)}{s^2} \frac{dt}{s} = \frac{X-Y}{s_0^3} \int_0^\infty \frac{(1+u)^2\big(A'(1+u)^2 + B'(1+u) + C'\big)}{v^2} \frac{du}{v} \,.
\end{equation}
The parameters $z_i$ are assumed to satisfy $\arg (z_i) < \pi$ for $1 \leq i \leq 4$, so that the integrals on both sides of equation~\eqref{eqn:int0} are well defined. The integral in equation~\eqref{eqn:int0} can be further decomposed as
\begin{equation}\label{eqn:int1}
	\mathcal{F} = \frac{X-Y}{s_0^3} \left( \mu_0 \int_0^\infty \frac{du}{v} - \sum_{i=1}^4 \mu_i \frac{(z_i - 1)^2}{\prod_{i \neq j}(z_i - z_j) } \int_0^\infty \frac{1}{z_i + u} \frac{du}{v}  \right) \,,
\end{equation}
where
\begin{equation}
	\mu_i = \begin{cases}
		    	A' &\text{for} \ i = 0 \,, \\
		    	A' z_i^2 - (2A' + B')z_i + (A' + B' + C') &\text{for} \ 1 \leq i \leq 4 \,.
		    \end{cases}
\end{equation}

Consider the \textit{symmetric integral} defined in \cites{gradshteyn2014table} as
\begin{equation}
	R_\mathcal{F} = \frac{1}{2} \int_0^\infty \frac{du}{\sqrt{(z_1 + u)(z_2 + u)(z_3 + u)(z_4 + u)}} \,.
\end{equation}
The derivatives of the symmetric integral with respect to parameters $z_i$ are given by
\begin{equation}
	\frac{\partial}{\partial z_i} R_\mathcal{F}(z_1,z_2,z_3,z_4) = -\frac{1}{4} \int_0^\infty \frac{\partial_{z_i}(v^2)}{v^2} \frac{du}{v} = -\frac{1}{4} \int_0^\infty \frac{1}{z_i + u} \frac{du}{v} \,.
\end{equation}
Therefore, the computation of the integral in equation~\eqref{eqn:int1} is reduced to the computation of a standard symmetric integral and its partial derivatives
\begin{equation}
	\mathcal{F} = \frac{2(X-Y)}{s_0^3} \left( \mu_0 + \sum_{i=1}^4 2\mu_i \frac{(z_i - 1)^2}{\prod_{i \neq j}(z_i - z_j) }\frac{\partial}{\partial z_i} \right) R_\mathcal{F} \,.
\end{equation}
Moreover, without loss of any generality, it is assumed that $X = 1$ and $Y = 0$.
\bigbreak
Thus, in summary, to evaluate the elliptic integral of the form
\begin{equation}
	\mathcal{F} = \int_0^1 \frac{A t^2 + B t  +  C  }{\Big( D t^4 + E t^3 + F t^2 + G t +  H \Big)^{\frac{3}{2}} }dt \,,
\end{equation}
it is enough to compute
\begin{equation}
	\mathcal{F} = \frac{2}{s_0^3} \left( \mu_0 + \sum_{i=1}^4 2\mu_i \frac{(z_i - 1)^2}{\prod_{i \neq j}(z_i - z_j) }\frac{\partial}{\partial z_i} \right) R_\mathcal{F} \,,
\end{equation}
with
\begin{equation}\label{eqn:symm}
	R_\mathcal{F} = \frac{1}{2} \int_0^\infty \frac{du}{\sqrt{(z_1 + u)(z_2 + u)(z_3 + u)(z_4 + u)}}
\end{equation}
and
\begin{equation}
	\mu_i = \begin{cases}
		    	A' &\text{for} \ i = 0 \,,\\
		    	A' z_i^2 - (2A' + B')z_i + (A' + B' + C') &\text{for} \ 1 \leq i \leq 4 \,,
		    \end{cases}
\end{equation}
where the parameters $z_i$ with $1 \leq i \leq 4$ are roots of the function
\begin{equation}
	z^4 - s_1 z^3 + s_2 z^2 - s_3 z + s_4 = 0 \,,
\end{equation}
with coefficients
\begin{equation}
	\begin{split}
	s_1 = \frac{E+2F+3G+4H}{D+E+F+G+H} \,, \ 
	s_2 = \frac{F+3G+6H}{D+E+F+G+H} \,,\\
	s_3 = \frac{G+4H}{D+E+F+G+H} \,,\
	s_4 = \frac{H}{D+E+F+G+H} \,,
	\end{split}
\end{equation}
that correspond to the elementary symmetric polynomials of the roots
\begin{equation}
 	s_1 = \sum_{1\le i \le 4} z_i \,, \  \ s_2 = \!\!\!\!\!\!\sum_{1\le i < j \le4}\!\!\! z_i z_j
 	\,, \  s_3 = \!\!\!\!\!\!\sum_{1\le i < j < k\le4}\!\!\!\!\!\! z_i z_j z_k
	 \,, \  s_4 = z_1 z_2 z_3 z_4 \,.
\end{equation}

The complexity of the original elliptic integral, $\mathcal{F}$, has been reduced by describing it in terms of the symmetric integral, $R_\mathcal{F}$, and its derivatives. However, as it stands, the evaluation of the symmetric integral is still a difficult process. The next step is to further decompose the problem by describing the symmetric integral in a more-applicable manner.

\subsection{Genus-One Curves and Their Jacobians}
Consider the Jacobian variety of the smooth genus-one curve $\mathcal{C}$ in equation~\eqref{eqn:genus_one}, $\text{Jac}(\mathcal{C})$. It is an elliptic curve, $\mathcal{E}$, that can be represented, over $\mathbb{C}$, as a fully-factorized, plane cubic curve with affine coordinates $(\xi,\eta) \in \mathbb{C}^2$ of the form
\begin{equation}
	\mathcal{E} \cong \text{Jac}(\mathcal{C}) : \quad \eta^2 = (K^2_2 + \xi)(K^2_3 + \xi)(K^2_4 + \xi) \,.
\end{equation}
Since the period described by the symmetric integral in equation~\eqref{eqn:symm} is a characteristic quantity of the Jacobian variety, $\text{Jac}(\mathcal{C})$, rather than the genus-one curve, $\mathcal{C}$, it is possible to reduce the symmetric integral, $R_\mathcal{F}$, to a period integral for the elliptic curve $\mathcal{E}$ \cites{carlson1997elliptic}.

Assume that the discriminant of $\mathcal{C}$ never vanishes (i.e. $\Delta_\mathcal{C} \neq 0$). Then, setting $z_i = Z_i^2$ for $1 \leq i \leq 4$, where the parameters $Z_i$ are located in the extended, open, right-half complex plane (i.e. $Z_1, ... , Z_4 \in \mathbb{C}_{\text{Re}>0} \cup \{ 0\}$), and defining the new parameters
\begin{equation}\label{eqn:parameters_EC}
	K_j = Z_1Z_j + Z_kZ_l \quad \text{for} \, \{j,k,l \} = \{2,3,4 \}, \{3,2,4 \}, \{4,2,3 \} \,,
\end{equation}
new rational functions $\xi, \eta \in \mathbb{C}(u,v)$ are defined on $\mathcal{C}$ by the expressions
\begin{align}\label{eqn:transfo2}
	 \xi  & =  2 v + 2 u^2 + (Z_1^2 + Z_2^2 +Z_3^2 +Z_4^2) u - 2 Z_1 Z_2 Z_3 Z_4 \,,\\
	 \eta&=  4u^3 +4 uv + (Z_1^2 + Z_2^2 +Z_3^2 +Z_4^2) v 
	 + (Z_1^2 Z_2^2 Z_3^2 + Z_1^2 Z_2^2 Z_4^2 + Z_1^2 Z_3^2 Z_4^2 + Z_2^2 Z_3^2 Z_4^2) \nonumber\\
	 & + 2 (Z_1^2Z_2^2+Z_1^2 Z_3^2 + Z_1^2 Z_4^2 + Z_2^2 Z_3^2 +Z_2^2 Z_4^2 + Z_3^2 Z_4^2) u + 3 (Z_1^2 + Z_2^2 +Z_3^2 +Z_4^2)  u^2 \,.\nonumber
\end{align}
It is easily verified that for $(u,v) = (0,0)$ and $(u,v) = (\infty,\infty)$, these rational functions return $(\xi,\eta) = (0,0)$ and $(\xi,\eta) = (\infty,\infty)$, respectively. In fact, the map $(u,v) \mapsto (\xi,\eta)$ defines a rational double cover, $\Phi : \mathcal{C} \dasharrow \mathcal{E}$, between the genus-one curve $\mathcal{C}$ and the elliptic curve $\mathcal{E}$. Furthermore, equation~\eqref{eqn:transfo2} induces an isomorphism $\text{Jac}(\mathcal{C}) \cong \mathcal{E}$, because $\mathcal{C}_{\bar{k}} \cong \mathcal{E}_{\bar{k}}$ over the algebraic closure $\bar{k}$, where $k$ is the function field of the moduli space. It readily follows that
\begin{equation}
	 \Phi^*\left(\frac{d\xi}{\eta}\right) = \frac{du}{v} \,.
\end{equation}

Based on the relation between the genus-one curve, $\mathcal{C}$, and its Jacobian variety, $\text{Jac}(\mathcal{C})$, the symmetric integral in equation~\eqref{eqn:symm} can be rewritten
\begin{equation}\label{eqn:symm2}
	\begin{split}
		R_\mathcal{F} &= \frac{1}{2} \int_0^\infty \frac{du}{\sqrt{(Z_1^2 + u)(Z_2^2 + u)(Z_3^2 + u)(Z_4^2 + u)}} \\ &= \frac{1}{2} \int_0^\infty \frac{d\xi}{\sqrt{(K_2^2 + \xi)(K_3^2 + \xi)(K_4^2 + \xi)}} \,,
	\end{split}
\end{equation}
where the relations between $Z_i$ and $K_i$ are defined by equation~\eqref{eqn:parameters_EC} \cites{carlson1997elliptic}. In this form, the symmetric integral can be evaluated using hypergeometric functions, leading to a closed-form solution to equation~\eqref{eqn:original_integral}.

\subsection{Relation to Hypergeometric Functions}
The generalized hypergeometric function $F_1$, or \textit{(first) Appell function}, is a formal extension of the Gauss hypergeometric function to two variables. For complex variables $x_1, x_2$ with $\max{(|x_1|, |x_2|)}<1$, and rational parameters $\alpha, \beta_1, \beta_2 \in (0,1) \cap \mathbb{Q}$ and $\gamma \in (0,1] \cap \mathbb{Q}$, it has the absolutely convergent power series expansion given by
\begin{equation}\label{eqn:series}
	\app{1}{\alpha; \beta_2, \beta_2 }{\gamma}{x_1, x_2} = \sum_{m=0}^\infty \sum_{n=0}^\infty  \frac{(\alpha)_{m+n} (\beta_1)_m (\beta_2)_n}{(\gamma)_{m+n}  m! n!} \, x_1^m x_2^n \,,
\end{equation}
where $(q)_k = \Gamma(q + k)/\Gamma(q)$ is the Pochhammer symbol for the rising factorial. Moreover, the function $F_1$ has an integral representation, for $\text{Re}(\alpha)>0$ and $\text{Re}(\gamma-\alpha)>0$, given by
\begin{equation}\label{eqn:integral}
 	\app{1}{\alpha; \beta_1, \beta_2}{\gamma}{x_1, x_2} = \frac{\Gamma(\gamma)}{\Gamma(\alpha)\Gamma(\gamma-\alpha)}
 	\int_0^1 \frac{w^{\alpha-1} (1-w)^{\gamma-\alpha-1}}{(1-x_1 w)^{\beta_1} (1-x_2 w)^{\beta_2}} \, dw \,.
\end{equation}
Restricting $x_2=0$, the classical Gauss hypergeometric function is regained,
\begin{equation}\label{eqn:2F1}
	  \app{1}{\alpha; \beta_1, \beta_2}{\gamma}{x_1, 0}   =\hpg{2}{1}{\alpha, \beta_1}{\gamma}{x_1} 
 	 = \sum_{m=0}^\infty \frac{(\alpha)_{m} (\beta_1)_m}{(\gamma)_{m}  m!} \, x_1^m  \,.
\end{equation}
If $\gamma-\alpha=1$, the Gauss hypergeometric function satisfies an important reduction identity---useful to reduce its defining parameters---given by
\begin{equation}\label{eqn:relat}
 	\hpg{2}{1}{\alpha+1, \beta+1}{\alpha+2}{x} = \frac{\alpha+1}{\beta x} \left( \frac{1}{(1-x)^\beta} -  \hpg{2}{1}{\alpha, \beta}{\alpha+1}{x} \right) \,.
\end{equation}
Using the integral representation~\eqref{eqn:integral}, the derivatives of $F_1$ are immediately found to be
\begin{multline}\label{eqn:F1}
	 \frac{\partial^{m+n}}{\partial x_1^m \partial x_2^n}  \app{1}{\alpha; \beta_1, \beta_2}{\gamma}{x_1, x_2}  \\ =\frac{(\alpha)_{m+n} (\beta_1)_m (\beta_2)_n}{(\gamma)_{m+n}} \, \app{1}{\alpha+m+n; \beta_1+m, \beta_2+n}{\gamma+m+n}{x_1, x_2}   \,.
\end{multline}

A relation is then found between the symmetric integral from equation~\eqref{eqn:symm2} and these hypergeometric functions as follows: using the transformation
\begin{equation}\label{eqn:vars}
 	x_1 = 1 - \frac{K_2^2}{K_4^2} \,,\; x_2= 1 - \frac{K_3^2}{K_4^2} \,, \;  w = \frac{K_4^2}{ K_4^2 + \xi} \,, \; y = - \frac{K_4 \eta}{(K_4^2 + \xi)^2}  \,,
\end{equation}
and the parameters $\alpha=\beta_1=\beta_2=\frac{1}{2}$ and $\gamma=\alpha+1$, the symmetric integral, $R_\mathcal{F}$ in equation~\eqref{eqn:symm2}, is rewritten in terms of $F_1$, creating the identity
\begin{equation}\label{eqn:symm_int}
	 R_\mathcal{F}   =  \frac{1}{K_4}  \app{1}{\frac{1}{2}; \frac{1}{2}, \frac{1}{2}}{\frac{3}{2}}{1 - \frac{K_2^2}{K_4^2}, 1 - \frac{K_3^2}{K_4^2}}  ,
\end{equation}
where $\Gamma(\frac{3}{2})/\Gamma(\frac{1}{2}) = \frac{1}{2}$, or, equivalently,
\begin{equation}\label{eqn:AppellF1}
	 R_\mathcal{F} = \frac{1}{Z_1Z_4+Z_2Z_3}  \app{1}{\frac{1}{2}; \frac{1}{2}, \frac{1}{2}}{\frac{3}{2}}{1 - \frac{(Z_1Z_2+Z_3Z_4)^2}{(Z_1Z_4+Z_2Z_3)^2}, 1 - \frac{(Z_1Z_3+Z_2Z_4)^2}{(Z_1Z_4+Z_2Z_3)^2}} \,,
\end{equation}
with derivatives found using equation~\eqref{eqn:F1}
\begin{multline}\label{eqn:AppellF1_derivatives}
	 \frac{\partial}{\partial Z_i} R_\mathcal{F} = \frac{\partial}{\partial Z_i} \left( \frac{1}{Z_1Z_4+Z_2Z_3} \right) \app{1}{\frac{1}{2}; \frac{1}{2}, \frac{1}{2}}{\frac{3}{2}}{x_1, x_2} \\
	 + \frac{1}{Z_1Z_4+Z_2Z_3} \left( \frac{1}{6} \, \app{1}{\frac{3}{2}; \frac{3}{2}, \frac{1}{2}}{\frac{5}{2}}{x_1, x_2}   \frac{\partial x_1}{\partial Z_i}  
	 +  \frac{1}{6} \, \app{1}{\frac{3}{2}; \frac{1}{2}, \frac{3}{2}}{\frac{5}{2}}{x_1, x_2}   \frac{\partial x_2}{\partial Z_i} \right) \,.
\end{multline}

These relations finally bring the original elliptic integral, from equation~\eqref{eqn:original_integral}, into a form that can be evaluated for the general case. Alternatively, any elliptic integral can be brought into one of three Legendre's canonical forms, usually denoted by
\begin{equation}
	 \mathrm{F}[\phi,k] = \int_0^\phi \frac{d\theta}{\Delta} \,, \quad  \mathrm{E}[\phi,k] = \int_0^\phi \Delta \, d\theta \,, \quad
	 \Pi[\phi, k, n] = \int_0^\phi \frac{d\theta}{\Delta (1+ n \sin^2{\theta})}
\end{equation} 
with $\Delta=\sqrt{1-k^2 \sin^2{\theta}}$. They are also called the incomplete elliptic integrals of the first, second, and third kind, respectively. The complete elliptic integrals are then recovered by
\begin{equation}
 	 \mathrm{K}(k) =   \mathrm{F}\left[\frac{\pi}{2},k\right] \,, \quad   \mathrm{E}(k) =   \mathrm{E}\left[\frac{\pi}{2},k\right] \,, \quad \Pi(k, n) = \Pi\left[\frac{\pi}{2},k, n\right] \,.
\end{equation} 
Substituting $\sin^2{\theta} = \frac{K_4^2-K_2^2}{K_4^2+\xi}$ into equation~\eqref{eqn:symm_int} yields
\begin{equation}
 	R_\mathcal{F} = \frac{1}{\sqrt{K_4^2-K_2^2}} \,   \mathrm{F}\left[ \sqrt{\frac{K_4^2-K_2^2}{K_4^2}}, \ \sqrt{\frac{K_4^2-K_2^2}{K_4^2- K_3^2}} \right]  \,.
\end{equation}
Expressions of this kind have been known in the literature; see \cites{carlson1997elliptic}. However, the use of the multivariate hypergeometric function $F_1$ is far superior as it pertains to its implementation, and the analysis hereafter is based on equation~\eqref{eqn:AppellF1}.  For example, the use of equation~\eqref{eqn:series} naturally allows a perturbation expansion of the solution to the general, parabolic vortex segment.

\subsection{Perturbation Expansion and Pencil of Elliptic Curves}\label{s:Perturbation}
It is heretofore shown that the general elliptic integral 
\begin{equation}\label{eqn:start_2}
	\mathcal{F}  =  \int_0^1 \frac{A t^2 + B t  +  C  }{\Big( D t^4 + E t^3 + F t^2 + G t +  H \Big)^{\frac{3}{2}} }dt
\end{equation}  
is computed by the expression of the form
\begin{equation}\label{eqn:goal_2}
 	 \mathcal{F} = \frac{2(Z_1 Z_2 Z_3 Z_4)^3}{H^{\frac{3}{2}}} \left( \mu_0+  \sum_{i=1}^4 \mu_{i}  \,  \frac{(Z^2_i-1)^2}{ Z_i \prod_{j\not = i}(Z^2_i-Z^2_j)} \frac{\partial}{\partial Z_i}\right)  \, R_\mathcal{F} \,,
\end{equation}   
where
\begin{equation}\label{eqn:mu_2}
	\mu_i = \begin{cases}
		    	A + B +C &\text{for} \ i = 0 \,,\\
		    	 (A + B+C) Z_i^4- (B+2C) Z_i^2+ C &\text{for} \ 1 \leq i \leq 4 \,,
		    \end{cases}
\end{equation}
and  $R_\mathcal{F}$ is the symmetric integral, that can be expressed in terms of the multivariate Appell hypergeometric function
\begin{equation}
 	R_\mathcal{F} = \frac{1}{Z_1Z_4+Z_2Z_3}  \app{1}{\frac{1}{2}; \frac{1}{2}, \frac{1}{2}}{\frac{3}{2}}{1 - \frac{(Z_1Z_2+Z_3Z_4)^2}{(Z_1Z_4+Z_2Z_3)^2}, 1 - \frac{(Z_1Z_3+Z_2Z_4)^2}{(Z_1Z_4+Z_2Z_3)^2}} \,,
\end{equation} 
whose derivatives, $\frac{\partial}{\partial Z_i}  \, R_\mathcal{F} $, are described by equation~\eqref{eqn:AppellF1_derivatives}. The parameters $Z_i^2$ for $1\le i \le 4$ are the roots of the function
\begin{equation}
	\big(Z^2\big)^4 - s_1 \big(Z^2\big)^3 + s_2 \big(Z^2\big)^2 - s_3 \big(Z^2\big) + s_4 = 0 \,,
\end{equation}
with coefficients
\begin{equation}
	\begin{split}
	s_1 = \frac{E+2F+3G+4H}{D+E+F+G+H} \,, \ 
	s_2 = \frac{F+3G+6H}{D+E+F+G+H} \,,\\
	s_3 = \frac{G+4H}{D+E+F+G+H} \,,\
	s_4 = \frac{H}{D+E+F+G+H} \,,
	\end{split}
\end{equation}
that correspond to the elementary symmetric polynomials
\begin{equation}\label{eqn:sym_poly_Zi}
 	s_1 = \sum_{1\le i \le 4} Z_i^2 \,, \  \ s_2 = \!\!\!\!\!\!\sum_{1\le i < j \le4}\!\!\! Z_i^2 Z_j^2
 	\,, \  s_3 = \!\!\!\!\!\!\sum_{1\le i < j < k\le4}\!\!\!\!\!\! Z_i^2 Z_j^2 Z_k^2
	 \,, \  s_4 = Z_1^2 Z_2^2 Z_3^2 Z_4^2 \,.
\end{equation}
Equations~\eqref{eqn:goal_2} through~\eqref{eqn:sym_poly_Zi}, therefore, result in a closed-form solution to equation~\eqref{eqn:start_2} if the roots $Z_i^2$, with $1\le i \le 4$, are known. The roots can be be found point-wise as solutions of a fourth order polynomial equation. However, to ensure a differentiable behavior of the closed-form solution as these roots undergo branching, it is not enough to just consider such point-wise solutions; one also needs to describe the correct way of gluing these solutions. We will describe such a family of solutions over a small disc in the complex plane or equivalently, by a perturbation expansion.

Consider the value of the elliptic integral $\mathcal{F}$ for a pencil, $\mathcal{C}_{\epsilon}$, of genus-one curves varying over a complex disc of radius one (i.e. $\epsilon \in \mathbb{C}$ with $|\epsilon|<1$) having a singularity at $\epsilon=0$. Roughly speaking, the desired pencil, $\mathcal{C}_\epsilon$, whose period integral interpolates between the integral for a linear vortex element and a parabolic one, is obtained by setting 
\begin{equation}
	\begin{split}
 	Z_3 = 1 + \epsilon \zeta_3 \,,\\
 	Z_4 = 1 + \epsilon \zeta_4 \,,
 	\end{split}
 \end{equation}
such that these roots $Z_3^2$ and $Z_4^2$ coincide, and equation~\eqref{eqn:genus_one} becomes singular, for $\epsilon = 0$  where $\zeta_3, \zeta_4$ are complex numbers that will be determined presently. Thus, we have
\begin{equation}
	\mathcal{C}_\epsilon: \quad v^2 = \big(Z_1^2 + u\big)\big(Z_2^2 + u\big)\big((1 + \epsilon \zeta_3)^2 + u\big)\big((1 + \epsilon \zeta_4)^2 + u\big) \,,
\end{equation}
where it is also assumed that the roots for $\epsilon = 1$ are pairwise different, and satisfy $Z_1, ... , Z_4 \in \mathbb{C}_{\text{Re}>0} \cup \{ 0\}$. It is easily verified that the corresponding pencil of elliptic curves, $\mathcal{E}_\epsilon \cong \text{Jac}(\mathcal{C}_\epsilon)$, is a smooth family for $0 < |\epsilon | < 1$, and has an isolated singular fiber of Kodaira type $I_2$ at $\epsilon = 0$ \cites{kodaira1960compact1, kodaira1963compact2, kodaira1963compact3}. Using this pencil, a perturbation expansion for $\mathcal{F}$ is computed, with the form
\begin{equation}\label{eqn:perturbation1}
	 \mathcal{F}  = \mathcal{F}^{(0)} + \epsilon  \, \mathcal{F}^{(1)} + O(\epsilon^2) \,.
\end{equation} 

Reflect back to the parameterization of the parabolic vortex filament defined in equation~\eqref{e:conic_segment_function}. Defining, 
\begin{equation}\label{eqn:eps}
	\epsilon = |\boldsymbol{r}_0 - \boldsymbol{r}_1 - \boldsymbol{f}_0'| \,,
\end{equation}
the parabolic filament can be treated as a first-order perturbation of the linear vortex filament described in equation~\eqref{e:linear_segment_function}
\begin{equation}
	\boldsymbol{f}(t) = \epsilon \boldsymbol{\hat{r}} t^2 + \boldsymbol{f}_0't + \boldsymbol{f}_0  \,,
\end{equation}
where $\boldsymbol{\hat{r}}$ is the unit vector
\begin{equation}
	\boldsymbol{\hat{r}} = \frac{\boldsymbol{r}_0 - \boldsymbol{r}_1 - \boldsymbol{f}_0'}{|\boldsymbol{r}_0 - \boldsymbol{r}_1 - \boldsymbol{f}_0'|} \,,
\end{equation}
such that $\boldsymbol{f}_0' = (\boldsymbol{r}_0 - \boldsymbol{r}_1) - \epsilon \boldsymbol{\hat{r}}$.
With such an expansion in $\epsilon$, the asymptotic behavior of the polynomial coefficients in equation~\eqref{eqn:start_2} turns out to be
\begin{equation*}
	A(\epsilon) = \epsilon (\boldsymbol{f}_0' \times \boldsymbol{\hat{r}}) + O(\epsilon^2) ,\quad B(\epsilon) = 2 \epsilon (\boldsymbol{\hat{r}} \times \boldsymbol{r}_0) + O(\epsilon^2),\quad C(\epsilon) = \boldsymbol{f}_0' \times \boldsymbol{r}_0 + O(\epsilon^2) ,
\end{equation*}
\begin{equation}\label{eqn:params}
	D = \epsilon^2 + O(\epsilon^3) ,\quad E = 2 \epsilon (\boldsymbol{f}_0' \cdot \boldsymbol{\hat{r}}) + O(\epsilon^2) ,\quad 
	F = f_0'^2 - 2 \epsilon (\boldsymbol{r}_0 \cdot \boldsymbol{\hat{r}}) + O(\epsilon^2) ,
\end{equation}
\begin{equation*}
	G = -2\boldsymbol{f}_0' \cdot \boldsymbol{r}_0  + O(\epsilon^2)  ,\quad H = r_0^2 + O(\epsilon^2) \,.
\end{equation*}
The integral in equation~\eqref{eqn:start_2} can then be organized in terms of the integrals
\begin{equation}
	\mathcal{F}_i  =  \int_0^1 \frac{t^i }{\Big( D t^4 + E t^3 + F t^2 + G t +  H \Big)^{\frac{3}{2}} }dt \quad \text{for} \,\, 0 \leq i \leq 2 \,,
\end{equation} 
each with a perturbation expansion of the form
\begin{equation}
	\mathcal{F}_i  =  \mathcal{F}_i^{(0)}  + \, \epsilon \, \mathcal{F}_i^{(1)}  + O(\epsilon^2) \,.
\end{equation} 
Therefore, the perturbation expansion in equation~\eqref{eqn:perturbation1} can be further decomposed as
\begin{equation}\label{eqn:perturbation2}
	\mathcal{F}  =  \underbrace{C(0) \, \mathcal{F}_0^{(0)}}_{\mathcal{F}^{(0)}}  + \; \epsilon  \underbrace{\bigg(C(0) \,  \mathcal{F}_0^{(1)}  + B'(0) \, \mathcal{F}_1^{(0)} + A'(0) \, \mathcal{F}_2^{(0)} \bigg)}_{\mathcal{F}^{(1)}} +\, O(\epsilon^2) 
\end{equation} 
with $A'(0) = \frac{dA}{d\epsilon} \vert_{\epsilon=0}$ and $B'(0) = \frac{dB}{d\epsilon} \vert_{\epsilon=0}$.

In the case that $\epsilon =0$, the coefficients in equation~\eqref{eqn:params} can be seen to match those of the linear vortex case described by equation~\eqref{e:linear_integral_form}. Thus, when determining $\mathcal{F}^{(0)}$ and $\mathcal{F}^{(1)}$ from equation~\eqref{eqn:perturbation1}, $\mathcal{F}^{(0)}$ is by construction the solution to the influence of a linear vortex filament, equation~\eqref{e:linear_segment_V}, and $\mathcal{F}^{(1)}$ is a first-order approximation of the parabolic effects. To start, consider a pencil 
\begin{equation}\label{eqn:pencil_roots}
	\mathcal{C}_\epsilon: \quad v^2 = \big(Z_1(\epsilon)^2 + u\big)\big(Z_2(\epsilon)^2 + u\big)\big(Z_3(\epsilon)^2 + u\big)\big(Z_4(\epsilon)^2 + u\big) \,,
\end{equation}
with the roots, $Z_i = Z_i(\epsilon)$ for $1 \leq i \leq 4$,  given by
\begin{equation}\label{eqn:roots}
	\begin{split}
	Z_1 &=   \frac{\sqrt{\boldsymbol{r}_0 \cdot \boldsymbol{r}_1 + r_0 r_1} + \sqrt{\boldsymbol{r}_0 \cdot \boldsymbol{r}_1 - r_0 r_1}}{\sqrt{2} r_1} - \epsilon  \frac{\sqrt{2}}{2}\frac{Z^{(1)}_N}{Z_D}\,,\\ 
	Z_2 &=   \frac{\sqrt{\boldsymbol{r}_0 \cdot \boldsymbol{r}_1 + r_0 r_1} - \sqrt{\boldsymbol{r}_0 \cdot \boldsymbol{r}_1 - r_0 r_1}}{\sqrt{2} r_1} + \epsilon  \frac{\sqrt{2}}{2}\frac{Z^{(2)}_N}{r_1^3 Z_D} \,, \\ 
	Z_3 &=  1 + \frac{\epsilon}{2} \frac{\boldsymbol{f}'_0 \cdot \boldsymbol{\hat{r}} + \sqrt{2(\boldsymbol{r}_0 \cdot \boldsymbol{r}_1) + (\boldsymbol{f}_0' \cdot \boldsymbol{\hat{r}})^2 - r_0^2 - r_1^2}}{2(\boldsymbol{r}_0 \cdot \boldsymbol{r}_1) - r_0^2 - r_1^2} \,, \\ 
	Z_4 &=  1 - \frac{\epsilon}{2} \frac{1}{\boldsymbol{f}'_0 \cdot \boldsymbol{\hat{r}} + \sqrt{2(\boldsymbol{r}_0 \cdot \boldsymbol{r}_1) + (\boldsymbol{f}_0' \cdot \boldsymbol{\hat{r}})^2 - r_0^2 - r_1^2}} \,.
	\end{split}
\end{equation}
The quantities $Z_N^{(i)}$ and $Z_D$ with $i=1, 2$ in equation~\eqref{eqn:roots} are then obtained by fine-tuning the elliptic pencil so the resulting elliptic integral matches the asymptotic behavior given by equation~\eqref{eqn:params}, resulting in the expressions
\begin{equation*}
\scalemath{0.9}{
\begin{aligned}
Z^{(1)}_N = &  r_0^3 \left( (\boldsymbol{f}_0' \cdot \boldsymbol{\hat{r}}) \sqrt{(\boldsymbol{r}_0 \cdot \boldsymbol{r}_1)^2 - r_0^2 r_1^2} + (\boldsymbol{r}_1 \cdot \boldsymbol{\hat{r}}) r_0^2 + (\boldsymbol{r}_0 \cdot \boldsymbol{\hat{r}}) r_1^2 - (\boldsymbol{r}_0 \cdot \boldsymbol{r}_1) \Big((\boldsymbol{r}_0+\boldsymbol{r}_1) \cdot \boldsymbol{\hat{r}}\Big) \right)\,,\\
Z^{(2)}_N = &\bigg(  \Big(-r_0^4 r_1^2 + 4 (\boldsymbol{r}_0 \cdot \boldsymbol{r}_1)  r_0^2 r_1^2 + 4 (\boldsymbol{r}_0 \cdot \boldsymbol{r}_1)^2 r_0^2 - 8 (\boldsymbol{r}_0 \cdot \boldsymbol{r}_1)^3\Big) (\boldsymbol{f}_0' \cdot \boldsymbol{\hat{r}})\\
& - (4 \boldsymbol{r}_0 \cdot \boldsymbol{r}_1 - r_0^2 r_1^2) ( \boldsymbol{r}_0 - \boldsymbol{r}_1)^2 (\boldsymbol{r}_0 \cdot \boldsymbol{\hat{r}}) \bigg)
\sqrt{\boldsymbol{r}_0 \cdot \boldsymbol{r}_1 + r_0r_1} \sqrt{\boldsymbol{r}_0 \cdot \boldsymbol{r}_1-r_0r_1} \\
& + \bigg(r_0^4 r_1^4 + 3 (\boldsymbol{r}_0 \cdot \boldsymbol{r}_1)  r_0^4 r_1^2 - 8 (\boldsymbol{r}_0 \cdot \boldsymbol{r}_1)^2 r_0^2 r_1^2 - 4 (\boldsymbol{r}_0 \cdot \boldsymbol{r}_1)^3 r_0^2 + 8 (\boldsymbol{r}_0 \cdot \boldsymbol{r}_1)^4\bigg) (\boldsymbol{f}_0' \cdot \boldsymbol{\hat{r}})\\
& + \bigg( 4 (\boldsymbol{r}_0 \cdot \boldsymbol{r}_1)^2 - 3 r_0^2 r_1^2\bigg)  (\boldsymbol{r}_0 - \boldsymbol{r}_1)^2 (\boldsymbol{r}_0 \cdot \boldsymbol{r}_1)  (\boldsymbol{r}_0 \cdot \boldsymbol{\hat{r}}) \,, \\
Z_D = &   \big(2 \boldsymbol{r}_0 \cdot \boldsymbol{r}_1 - r_0^2 - r_1^2\big)
 \bigg( (\boldsymbol{r}_0 \cdot \boldsymbol{r}_1 - r_0 r_1) (2 \boldsymbol{r}_0 \cdot \boldsymbol{r}_1 + r_0 r_1) \sqrt{\boldsymbol{r}_0 \cdot \boldsymbol{r}_1 + r_0 r_1} \\
& -  (\boldsymbol{r}_0 \cdot \boldsymbol{r}_1 + r_0 r_1) (2 \boldsymbol{r}_0 \cdot \boldsymbol{r}_1 - r_0 r_1)\sqrt{\boldsymbol{r}_0 \cdot \boldsymbol{r}_1 - r_0 r_1}\bigg) \,.
\end{aligned}}
\end{equation*}
With these roots and using equations~\eqref{eqn:parameters_EC}, an expansion of equation~\eqref{eqn:vars} of the form $x_i = x_i^{(0)} + \epsilon x_i^{(1)} + O(\epsilon^2)$ for $1 \leq i \leq 2$, yields
\begin{equation}\label{eqn:x1x2_expansion}
	x_1^{(0)} = \frac{2\boldsymbol{r}_0 \cdot \boldsymbol{r}_1 - r_0^2 - r_1^2}{2(\boldsymbol{r}_0 \cdot \boldsymbol{r}_1 + r_0 r_1)}\,,  \qquad x_2^{(0)} = 0 \,.
 \end{equation}
This shows why expressing the elliptic integral in terms of the multivariate hypergeometric function $F_1$ is a powerful analytical tool: in the limit $\epsilon \to 0$ one finds $x_2=0$ and the multivariate hypergeometric function $F_1$ restricts to the Gauss hypergeometric function in equation~\eqref{eqn:2F1}.
 
Carrying out the series expansion of equation~\eqref{eqn:goal_2}, the leading order term, $\mathcal{F}^{(0)}$, is found to be
 \begin{equation}
 	\begin{split}
 	 \mathcal{F}^{(0)} & = \frac{2 C(0) \,  Z_1^2 Z_2^2 (1 + Z_1 Z_2)^2}{H^{\frac{3}{2}} (Z_1 + Z_2)^3} \app{1}{\frac{1}{2}; \frac{1}{2}, \frac{1}{2}}{\frac{3}{2}}{x^{(0)}_1, 0} \\
  	&  - \frac{4 C(0) \, Z_1^2 Z_2^2 (1 + Z_1 Z_2)^2 (1 - Z_1^2)(1 - Z_2^2)}{H^{\frac{3}{2}} (Z_1 + Z_2)^5} \frac{\partial}{\partial x_1}\app{1}{\frac{1}{2}; \frac{1}{2}, \frac{1}{2}}{\frac{3}{2}}{x^{(0)}_1, 0} + O(\epsilon) \,.
	 \end{split} 
 \end{equation}
Using equations~\eqref{eqn:2F1} -- \eqref{eqn:F1}, and equations~\eqref{eqn:roots} -- \eqref{eqn:x1x2_expansion}, this term can be rewritten to yield
 \begin{equation}
 	 \mathcal{F}^{(0)} = \frac{2 C(0) \, Z_1^2 Z_2^2 (1 + Z_1 Z_2)}{H^{\frac{3}{2}} (Z_1 + Z_2)^2} + O(\epsilon)  = \frac{C(0) \, (r_0 + r_1)}{ (\boldsymbol{r}_0 \cdot \boldsymbol{r}_1 + r_0 r_1)r_0 r_1} \,,
 \end{equation}
matching precisely equation~\eqref{e:linear_segment_V}, as designed. A similar computation is repeated for each of the remaining terms in equation~\eqref{eqn:perturbation2}, resulting in the set of expressions
\begin{equation}\label{eqn:explicit}
\scalemath{0.8}{
\begin{aligned}
 \mathcal{F}^{(0)}_0 & = \frac{r_0 + r_1}{r_0 r_1(\boldsymbol{r}_0 \cdot \boldsymbol{r}_1 + r_0 r_1 )}  \,,\\
 \mathcal{F}^{(1)}_0 & =  \frac{(\boldsymbol{r}_0 \cdot \boldsymbol{\hat{r}}) (\boldsymbol{r}_0 \cdot \boldsymbol{r}_1 + 2 r_0 r_1 + r_1^2)}{r_1^3 (\boldsymbol{r}_0 \cdot \boldsymbol{r}_1 + r_0r_1)^2} 
 - \frac{(\boldsymbol{f}_0' \cdot \boldsymbol{\hat{r}}) (\boldsymbol{r}_0 \cdot \boldsymbol{r}_1 + 2 r_0 r_1)}{r_1^3 (\boldsymbol{r}_0 \cdot \boldsymbol{r}_1 + r_0r_1)^2} \\
 &- \frac{ (r_0 + r_1)\sqrt{(\boldsymbol{f}_0' \cdot \boldsymbol{\hat{r}})^2 + 2 \boldsymbol{r}_0 \cdot \boldsymbol{r}_1 - r_0^2 - r_1^2}}{(2 \boldsymbol{r}_0 \cdot \boldsymbol{r}_1 - r_0^2 - r_1^2) \sqrt{\boldsymbol{r}_0 \cdot \boldsymbol{r}_1 - r_0 r_1} (\boldsymbol{r}_0 \cdot \boldsymbol{r}_1 + r_0r_1) ^{\frac{3}{2}}} \\
 &+ \frac{2 \sqrt{2}\sqrt{(\boldsymbol{f}_0' \cdot \boldsymbol{\hat{r}})^2+2\boldsymbol{r}_0 \cdot \boldsymbol{r}_1 - r_0^2 - r_1^2}}{ (2\boldsymbol{r}_0 \cdot \boldsymbol{r}_1-r_0^2-r_1^2)(\boldsymbol{r}_0 \cdot \boldsymbol{r}_1+r_0r_1)\sqrt{\boldsymbol{r}_0 \cdot \boldsymbol{r}_1-r_0r_1}}  
  \hpg{2}{1}{\frac{1}{2}, -\frac{1}{2}}{\frac{3}{2}}{  \frac{2\boldsymbol{r}_0 \cdot \boldsymbol{r}_1 - r_0^2-r_1^2}{2(\boldsymbol{r}_0 \cdot \boldsymbol{r}_1 +r_0r_1)}} \\
  & - \frac{\sqrt{2} \, \sqrt{(\boldsymbol{f}_0' \cdot \boldsymbol{\hat{r}})^2+2\boldsymbol{r}_0 \cdot \boldsymbol{r}_1 - r_0^2 - r_1^2}}{(2\boldsymbol{r}_0 \cdot \boldsymbol{r}_1-r_0^2-r_1^2)(\boldsymbol{r}_0 \cdot \boldsymbol{r}_1+r_0r_1)\sqrt{\boldsymbol{r}_0 \cdot \boldsymbol{r}_1-r_0r_1}}  
  \hpg{2}{1}{\frac{1}{2}, \frac{1}{2}}{\frac{3}{2}}{  \frac{2\boldsymbol{r}_0 \cdot \boldsymbol{r}_1 - r_0^2-r_1^2}{2(\boldsymbol{r}_0 \cdot \boldsymbol{r}_1 +r_0r_1)}}   \,,\\
 \mathcal{F}^{(0)}_1 & = \frac{1}{r_1 (\boldsymbol{r}_0 \cdot \boldsymbol{r}_1 + r_0 r_1)} \,,\\
 \mathcal{F}^{(0)}_2 & = \frac{(2\boldsymbol{r}_0 \cdot \boldsymbol{r}_1 + r_0 (r_1- r_0) )}{r_1 (\boldsymbol{r}_0 \cdot \boldsymbol{r}_1 + r_0 r_1)(2\boldsymbol{r}_0 \cdot \boldsymbol{r}_1 - r_0^2 - r_1^2)} \\
 & - \frac{\sqrt{2}}{(2\boldsymbol{r}_0 \cdot \boldsymbol{r}_1-r_0^2-r_1^2)\sqrt{\boldsymbol{r}_0 \cdot \boldsymbol{r}_1+r_0r_1}}  
  \hpg{2}{1}{\frac{1}{2}, \frac{1}{2}}{\frac{3}{2}}{  \frac{2\boldsymbol{r}_0 \cdot \boldsymbol{r}_1 - r_0^2-r_1^2}{2(\boldsymbol{r}_0 \cdot \boldsymbol{r}_1 +r_0r_1)}}  \,,
\end{aligned}}
\end{equation}
that, when used with equation~\eqref{eqn:explicit} in equation~\eqref{eqn:perturbation2}, provides an approximation of equation~\eqref{eqn:start_2} up to $O(\epsilon^2)$.
\section{Validation and Comparison of Predictive Methods}
Consider five methods to evaluate the integral
\begin{equation}\label{e:parabolic_integral_form2}
	\boldsymbol{V} = \frac{\Gamma}{4 \pi} \int_0^{1} \frac{\Big( A t^2 + B t  +  C \Big) dt}{\Big( D t^4 + E t^3 + F t^2 + G t +  H \Big)^{\frac{3}{2}} } \,,
\end{equation}
where
\begin{equation*}
	A(\epsilon) = \boldsymbol{f}_0' \times (\boldsymbol{r}_0 - \boldsymbol{r}_1),\quad B(\epsilon) = -2(\boldsymbol{r}_1 + \boldsymbol{f}_0') \times \boldsymbol{r}_0,\quad C(\epsilon) = \boldsymbol{f}_0' \times \boldsymbol{r}_0,
\end{equation*}
\begin{equation*}
	D = (\boldsymbol{r}_0 - \boldsymbol{r}_1 - \boldsymbol{f}_0')^2 ,\quad E = 2(\boldsymbol{r}_0 - \boldsymbol{r}_1 - \boldsymbol{f}_0') \cdot \boldsymbol{f}_0',\quad F = f_0'^2 - 2\boldsymbol{r}_0 \cdot (\boldsymbol{r}_0 - \boldsymbol{r}_1 - \boldsymbol{f}_0'),
\end{equation*}
\begin{equation*}
	G = -2\boldsymbol{r}_0 \cdot \boldsymbol{f}_0' ,\quad H = r_0^2 \,.
\end{equation*}

First, in section~\ref{s:Integral}, equation~\eqref{e:parabolic_integral_form2} is related to hypergeometric functions, resulting in the closed-form expressions 
\begin{equation}\label{eqn:goal_3}
 	 \boldsymbol{V} = \frac{\Gamma}{4 \pi} \frac{2(Z_1 Z_2 Z_3 Z_4)^3}{H^{\frac{3}{2}}} \left( \mu_0+  \sum_{i=1}^4 \mu_{i}  \,  \frac{(Z^2_i-1)^2}{ Z_i \prod_{j\not = i}(Z^2_i-Z^2_j)} \frac{\partial}{\partial Z_i}\right)  \, R_\mathcal{F} \,,
\end{equation}   
where equations~\eqref{eqn:mu_2} -- \eqref{eqn:sym_poly_Zi} describe $\mu_i$, $R_\mathcal{F}$, and $Z_i$. equation~\eqref{eqn:goal_3} results in a closed-form solution to equation~\eqref{e:parabolic_integral_form2} if the roots $Z_i^2$, with $1\le i \le 4$, can be readily found.

Second, in section~\ref{s:Perturbation}, equation~\eqref{eqn:goal_3} is expanded in $\epsilon$, leading to the approximation
\begin{equation}\label{eqn:perturbation3}
	\boldsymbol{V} =  \frac{\Gamma}{4 \pi} \bigg( C(0) \, \mathcal{F}_0^{(0)}  + \epsilon \Big( C(0) \, \mathcal{F}_0^{(1)}  + B'(0) \, \mathcal{F}_1^{(0)} + A'(0)  \, \mathcal{F}_2^{(0)} \Big) +\, O(\epsilon^2) \bigg) \,,
\end{equation} 
where equations~\eqref{eqn:eps} and~\eqref{eqn:params} define $\epsilon$, $A$, $B$, and $C$, and equation~\eqref{eqn:explicit} defines $\mathcal{F}_i^{(j)}$. This expansion is about the case of a vortex line segment (i.e. $\epsilon = 0$), and thus the best results can be expected when the parabolic vortex segment has little curvature.

Third, discussed in section~\ref{s:Bliss_method}, is an approximation made in \cites{bliss1987new} for the case of the parabolic parameterization
\begin{align}\label{e:conic_segment_function2}
	\boldsymbol{f}(t) &= (\boldsymbol{r}_0 - \boldsymbol{r}_1 - \boldsymbol{f}_0') t^2 + \boldsymbol{f}_0't + \boldsymbol{f}_0 \nonumber \,,\\
	d\boldsymbol{f}(t) &= \big(2(\boldsymbol{r}_0 - \boldsymbol{r}_1 - \boldsymbol{f}_0')t + \boldsymbol{f}_0' \big)dt \,,
\end{align}
where
\begin{equation*}
	\boldsymbol{f}_0 = \big[ -\ell, \varepsilon \ell^2, 0\big] \,,
\end{equation*}
\begin{equation}\label{e:vortex_form_sym}
	\boldsymbol{r}_0 - \boldsymbol{r}_1 = \big[ 2 \ell, 0, 0 \big]\,,
\end{equation}
\begin{equation*}
	\boldsymbol{f}'_0 = \big[ 2 \ell, -4\varepsilon \ell^2 , 0 \big]\,,
\end{equation*}
 in which equation~\eqref{e:parabolic_integral_form2} is approximated by the integral
\begin{equation}\label{e:symmetric_parabolic_integral2}
\scalemath{0.9}{	\boldsymbol{V} = \frac{\Gamma}{4 \pi} \int_{-\ell}^{\ell} \frac{ \big[ 2\varepsilon z_0 x,\  -z_0, \ -2\varepsilon x x_0 + \varepsilon x^2 + y_0 \big]dx}{ \Big( (1 - 2\varepsilon y_0 + \varepsilon^2 F_2)x^2 + (-2x_0 + \varepsilon^2 F_1)x + (x_0^2 + y_0^2 + z_0^2 + \varepsilon^2 F_0) \Big)^{3/2}} \,,}
\end{equation}
where $F_2$, $F_1$, and $F_0$ must be tuned to accurately replicate the original integral. Note the difference between $\varepsilon$, used in \cites{bliss1987new} and defined in equation~\eqref{e:vareps}, and $\epsilon$,  from equation~\eqref{eqn:perturbation3} and defined in equation~\eqref{eqn:eps}.

Fourth, equation~\eqref{e:parabolic_integral_form2} can be evaluated numerically. Numerical integration can be an effective means of evaluating an integral, so long as the integral is subdivided into a sufficient number of sections, and the integral itself does not exhibit highly oscillatory or asymptotic behavior. In the case of equation~\eqref{e:parabolic_integral_form2}, the denominator tends to zero as the point at which the induced velocity is calculated moves close to the vortex filament, thus, accurate numerical results may be expected away from the vortex, but the results may lose fidelity as the vortex is approached.

Finally, the results of the first four methods can be compared to the approximation of the parabolic vortex segment by several of the straight vortex segments described in section~\ref{s:Linear_segment}. Such a comparison will provide insight into the number of straight segments  needed to accurately reproduce the parabolic segment, and perhaps hint at the advantages provided by the use of parabolic vortices.

The comparisons made in this work will be restricted to a planar problem in which the point of interest resides in the same plane as a parabolic vortex segment of the form 
\begin{equation*}
	\boldsymbol{f}_0 = \big[ -\ell, \varepsilon \ell^2, 0\big] \,,
\end{equation*}
\begin{equation}\label{e:vortex_form_asym}
	\boldsymbol{r}_0 - \boldsymbol{r}_1 = \big[ 2 \ell, 0, 0 \big] \,, 
\end{equation}
\begin{equation*}
	\boldsymbol{f}'_0 = \big[ 4 \kappa + 2 \ell, -4\varepsilon \ell^2 , 0 \big] \,,
\end{equation*}
which is a simple extension of equation~\eqref{e:vortex_form_sym} that allows for the description of asymmetric, parabolic vortices when $\kappa \neq 0$. Sample points will be taken along the $y$-axis ($x = 0$). Additionally, all comparisons will assume $\Gamma = 4\pi$, to further reduce the degrees of freedom under consideration.

\subsection{Analytic vs Numerical Integration}
\begin{figure}
	\centering
	\includegraphics{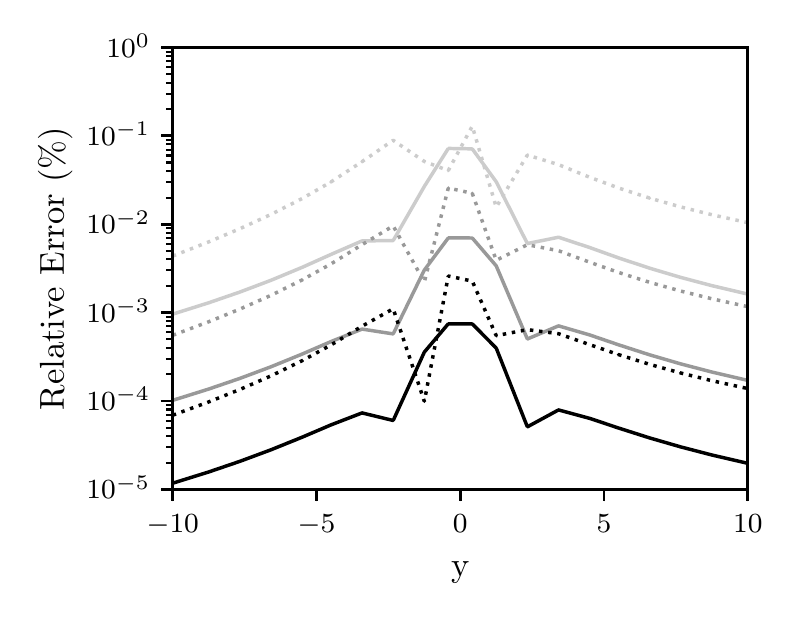}
	\caption{The relative error of the numerical integration with respect to the analytic solution expressed in equation~\eqref{eqn:goal_3}, for $\{\varepsilon, \kappa\} = \{-0.01, 0.0 \}$ (solid) and $\{\varepsilon, \kappa\} = \{-0.1, 0.5 \}$ (dashed). Each line represents a different number of intervals used in the numerical integration, $n = \{10, 20, 40 \}$ (lightest to darkest).}
	\label{f:AvN}
\end{figure}

To validate the analytic solution derived herein, equation~\eqref{e:parabolic_integral_form2} is numerically integrated, using Simpson's rule, and compared to the results of equation~\eqref{eqn:goal_3}. Numerical validation of this nature is manifest when the predictions obtained by numerical integration approach those obtained using the analytic formula as the number of intervals used in the integration increases towards infinity. The comparison between numerical integration and the analytic solution is made both for a symmetric case with little curvature, $\{\varepsilon, \kappa\} = \{-0.01, 0.0 \}$, and an asymmetric case with larger curvature, $\{\varepsilon, \kappa\} = \{-0.1, 0.5 \}$. Each of these cases is predicted using three different quantities of intervals, $n = \{10, 20, 40 \}$, to determine numerical convergence and identify the relationship between the number of intervals used and the accuracy of the prediction. The results are shown in figure~\ref{f:AvN}.

It can seen in figure~\ref{f:AvN} that the numerical integration is convergent for the two cases displayed, observing that the error diminishes linearly on a logarithmic scale as the number of intervals increases. And, as predicted, the error is higher close to the vortex ($y \approx 0$), and decreases further away. It is also observed that the relative error of the numerical integration is larger for the asymmetric, high-curvature case than for the symmetric, low-curvature case. This suggests that a higher number of intervals is required as the vortex segment becomes increasingly dissimilar to a linear segment.

\subsection{Analytic vs Approximation by Several Straight Segments}
\begin{figure}
	\centering
	\includegraphics{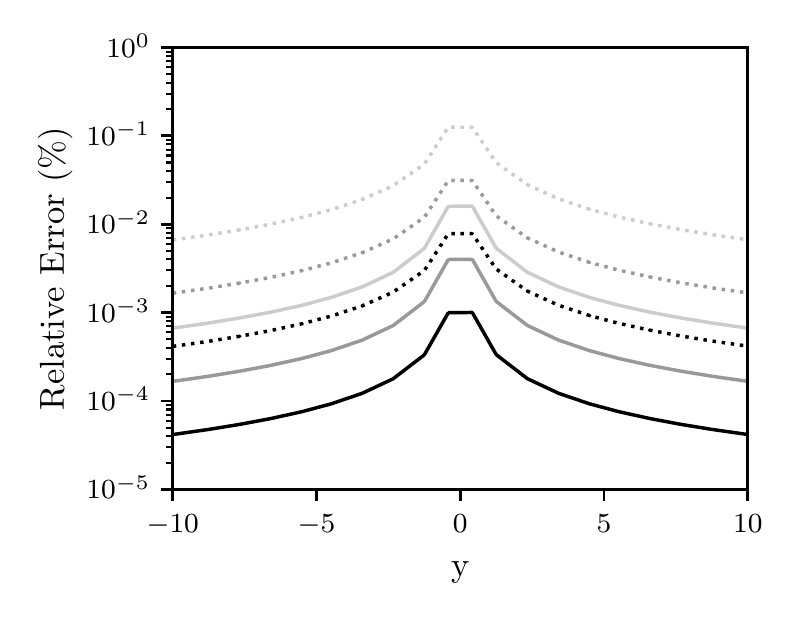}
	\caption{The relative error of the approximation of the vortex with several straight vortex segments with respect to the analytic solution expressed in equation~\eqref{eqn:goal_3}, for $\{\varepsilon, \kappa\} = \{-0.01, 0.0 \}$ (solid) and $\{\varepsilon, \kappa\} = \{-0.1, 0.5 \}$ (dashed). Each line represents a different number of linear segments used, $n = \{10, 20, 40 \}$ (lightest to darkest).}
	\label{f:AvL}
\end{figure}

To further corroborate the results obtained from the explicit formula in equation~\eqref{eqn:goal_3} and numerical integration, predictions are made by approximating the parabolic vortex segment with several linear vortex segments. Again, the comparison is made for a symmetric case with little curvature, $\{\varepsilon, \kappa\} = \{-0.01, 0.0 \}$, and an asymmetric case with larger curvature, $\{\varepsilon, \kappa\} = \{-0.1, 0.5 \}$, and each case is predicted using three quantities of linear vortex segments, $n = \{10, 20, 40 \}$. The results are shown in figure~\ref{f:AvL}.

As with the results from numerical integration, it can seen in figure~\ref{f:AvL} that the linear-segment approximation is convergent for the two cases displayed, and that the relative error is larger for the asymmetric, high-curvature case than for the symmetric, low-curvature case. These cases suggest that at least ten linear vortex segments are required to represent a single parabolic vortex segment within 1\%. It is noted that the relative error of the linear-vortex approximation is, in general, higher than that of numerical integration, for an equal number of intervals and segments, though the relative error of both methods is well under 1\% for the cases observed.

\subsection{Analytic vs Approximation in \texorpdfstring{\cite{bliss1987new}}{[5]} }
\begin{figure}
	\centering
	\includegraphics{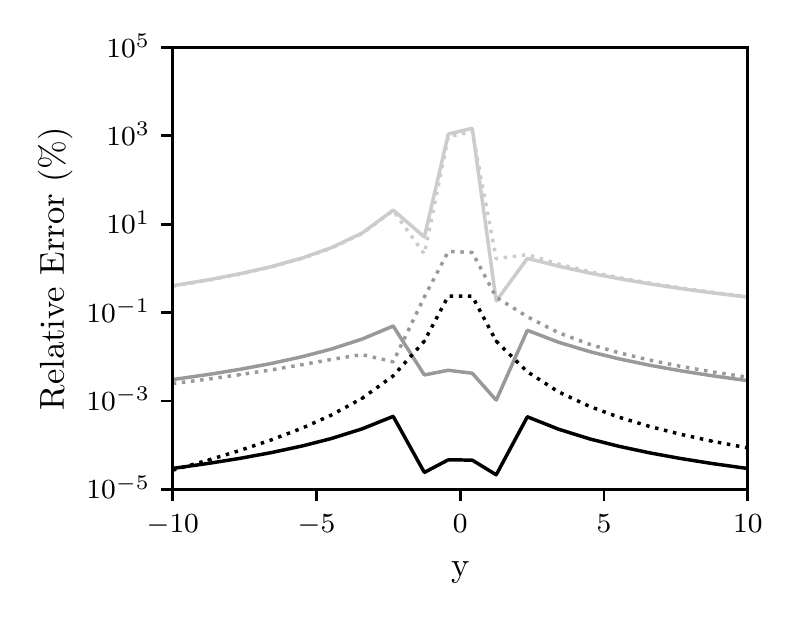}
	\caption{The relative error of the approximation in \cites{bliss1987new} in  equation~\eqref{e:symmetric_parabolic_integral2} with respect to the analytic solution expressed in equation~\eqref{eqn:goal_3}. Each line represents a different value of $\varepsilon$ and $\kappa$, where $\varepsilon = \{-0.01, -0.1, -1.0\}$(darkest to lightest) and $\kappa = \{0.0, 0.5\}$ (solid, dashed).}
	\label{f:AvB}
\end{figure}

Having validated the analytic solution using two computational methods, a comparison is performed between the approximation proposed in \cites{bliss1987new} in equation~\eqref{e:symmetric_parabolic_integral2}  and the explicit formula expressed in equation~\eqref{eqn:goal_3}. The comparison is made with three values of $\epsilon$ (from -0.01 to -1.0) for both a symmetric ($\kappa = 0.0$) and an asymmetric ($\kappa = 0.5$) parabolic vortex. The results are shown in figure~\ref{f:AvB}.

Again, the results of this comparison look as expected---the error of the approximation diminishes as $\epsilon$ tends to zero. The case of the symmetric vortex results in less error than the asymmetric vortex, close to the vortex itself ($y \approx 0$), through the relative error for both cases is similar far away from the vortex, suggesting that the effect of the asymmetry on the induced velocity far from the vortex is negligible. Similarly, as $\epsilon$ grows larger---such as the case that $\epsilon = -1.0$ in figure~\ref{f:AvB}---the effect of asymmetry on the relative error of the approximation also appears to become insignificant. This is perhaps because the error due to the large curvature overshadows any contribution to the error from the asymmetry.

\subsection{Analytic vs Perturbation Expansion}
\begin{figure}
	\centering
	\includegraphics{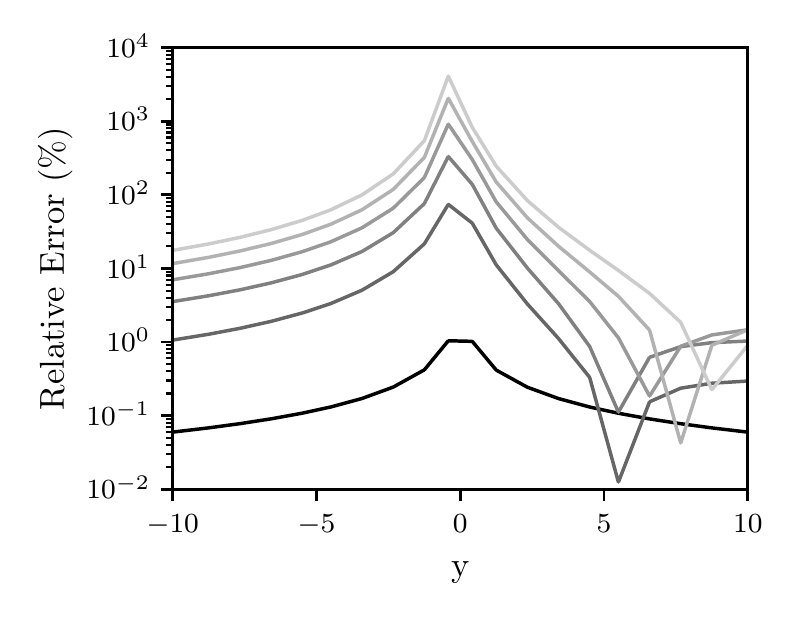}
	\caption{The relative error of the perturbation expansion in equation~\eqref{eqn:perturbation3} with respect to the analytic solution expressed in equation~\eqref{eqn:goal_3}. Each line represents a different value of $\epsilon$ (created by varying $\varepsilon$ and $\kappa$), such that $\epsilon = \{0.004, 0.4, 0.8, 1.2, 1.6, 2.0\}$ (darkest to lightest).}
	\label{f:AvP}
\end{figure}

It is now of interest to compare the analytic solution expressed in equation~\eqref{eqn:goal_3} with its perturbation expansion, expressed in equation~\eqref{eqn:perturbation3}. The comparison is made by simultaneously varying $\varepsilon$ from -0.001 to -0.09 and $\kappa$ from 0.0 to 0.5 in order to vary $\epsilon$ from 0.004 to 2.0, as defined by equation~\eqref{eqn:eps}. The results are shown in figure~\ref{f:AvP}.

As expected, the error of the perturbation expansion, relative to the analytical solution, decreases as $\epsilon$ approaches zero. Unfortunately, in order for the $O(\epsilon^2)$ expansion described in equation~\eqref{eqn:perturbation3} to approximate the analytical solution to within 1\%, $\epsilon$ must remain prohibitively small for engineering applications. The inclusion of more terms in the expansion would generalize the use of the approximation, but each term is increasingly more mathematically complex to derive.

\subsection{Computation Cost Comparison}
In addition to comparisons of accuracy, it is of practical interest to compare the computational cost of each prediction method.  Table~\ref{t:comp_cost} shows a comparison of the time required by the five methods discussed previously to predict the influence of the vortex at various distances along the $y$-axis. Each of the methods was naively implemented in Python, using intrinsic functions and libraries, and without substantial optimization for efficiency. The case wherein $\{\varepsilon, \kappa\} = \{-0.1, 0.5 \}$ was selected to more clearly demonstrate the increase in computation cost required to compute the analytic prediction near the vortex itself.
\begin{table}
  \begin{center}
\def~{\hphantom{0}}
  \begin{tabular}{cccccc}
      $y$  & $Analytic$   &   $Perturbation$ & Bliss \cite{bliss1987new} & $Numerical$ & $Linear$\\
      $(x = 0)$  & $(Eq. \eqref{eqn:goal_3})$ & $(Eq. \eqref{eqn:perturbation3})$ & $(Eq. \eqref{e:symmetric_parabolic_integral2})$ & $Integration$ & $Segments$ \\[3pt]
       -10.00   &3.90E-2  &8.53E-4  & 7.90E-5 & 2.21E-4 & 1.85E-4 \\
       ~-8.75   &4.05E-2  & 5.36E-4 & 7.60E-5 &  1.90E-4&  1.68E-4\\
       ~-7.67   &4.28E-2  &  5.50E-4&  7.70E-5&  2.23E-4&  1.99E-4\\
       ~-6.58   & 4.59E-2 &  5.58E-4&  7.90E-5& 2.39E-4 &  1.77E-4\\
       ~-5.50   & 4.98E-2 &  5.41E-4&  7.60E-5&  1.89E-4& 1.65E-4 \\
       ~-4.50   &  5.67E-2&  6.30E-4&  1.10E-4&  2.21E-4&  1.75E-4\\
       ~-3.42   &  6.96E-2&  5.48E-4&  8.50E-5&  1.86E-4&  1.72E-4\\
       ~-2.33   &  1.51E-1&  5.78E-4&  8.90E-5&  2.14E-4&  1.77E-4\\
       ~-1.25   &  2.50E+0&  8.03E-4&  1.68E-4&  2.26E-4& 1.80E-4 \\
       ~-0.42   &  1.15E+0&  5.00E-3&  1.65E-4&  2.11E-4&  1.75E-4\\
  \end{tabular}
  \caption{Time (in seconds) required to predict the influence of the parabolic vortex segment defined by $\{\varepsilon, \kappa\} = \{-0.1, 0.5 \}$, by various methods at several locations along the $y$-axis.}
  \label{t:comp_cost}
  \end{center}
\end{table}

From table~\ref{t:comp_cost}, it can be observed that, while resulting in the highest level of accuracy, the evaluation of the full analytic solution comes at a substantially higher computational cost than that of the other methods. This is due to the cost of computing the multivariate Appell hypergeometric function, and finding the roots of a quartic polynomial. The cost decreases drastically for the the perturbation expansion, where a hypergeometric function in only one variable is computed. The remaining three methods require roughly the same time to compute, though the approximation in \cites{bliss1987new} is the fastest in all cases. It should be remembered that, while requiring a lower computational cost, each approximation is limited to certain geometries of parabolic vortex. The approximation made in \cites{bliss1987new} is limited to symmetric parabolic arcs with little curvature, the perturbation expansion is likewise limited to vortex segments with little curvature, and the numerical integration and linear-segments approximation decrease in accuracy when predicting induced velocities close to the vortex itself.

\section{Conclusion}
It is remarkable that an explicit formula for the velocity field induced by a parabolic vortex segment can be determined using classical algebraic geometry, in particular period integrals for a holomorphic family of elliptic surfaces. The problem of finding the velocity profile surrounding vortices is ubiquitous, and it is a crucial ingredient relevant for fluid and aerodynamics, including the large topic of turbulence. It might also find its application in the large field of quantum fluids and gases, where vortices carry quantized circulation, and the velocity is identified with the gradient of the phase of the superfluid. The integral of course also describes the magnetic field induced by a parabolic current loop, so that the solution may find its application also in such "dual" circumstances.
The explicit formulas in this paper have been developed using multivariate hypergeometric functions to predict the velocity induced by a general parabolic vortex segment, summarized by equation~\eqref{eqn:goal_3}. This formula is based on the integral that results from the Biot-Savart law, and is derived by constructing a genus-one whose period integrals provide the solution to the induced velocity. Moving from the genus-one curve to the corresponding Jacobian elliptic curve, the resulting elliptic integral can be evaluated explicitly using multivariate special functions. The evaluation of the formula in equation~\eqref{eqn:goal_3}, though analytically explicit, requires finding the roots of a quartic polynomial and the implementation of the multivariate first Appell hypergeometric function, which complicates its practical implementation. Using the carefully crafted pencil of genus-one curves in equation~\eqref{eqn:pencil_roots}, the series expansion of the first Appell hypergeometric function was used to derive a quadratic perturbation expansion to interpolate between the linear and the parabolic vortex segment. Equation~\eqref{eqn:goal_3} is validated through comparison to the predictions resulting from computational methods---numerical integration and the approximation of the parabolic vortex by several linear vortex segments. Figures~\ref{f:AvN} and~\ref{f:AvL} show that those numerical methods converge to the analytic formula as the number of intervals or segments increase towards infinity, corroborating the validity of the explicit formula.

\end{document}